\documentclass{iopconfser}

\usepackage[numbers,square,sort&compress]{natbib}
\usepackage{lmodern}

\usepackage{amsmath, amsthm, mathrsfs, amssymb, amsfonts, mathtools}
\usepackage{graphicx}
\usepackage{hyperref}
\hypersetup{colorlinks=true, linkcolor=blue, citecolor=magenta, filecolor=magenta, urlcolor=cyan}

\def\diff{\mathrm{d}}

\usepackage{orcidlink}

\begin{document}

\title{Time-periodic solutions of the conformally invariant wave equation on the Einstein cylinder}

\author{Filip Ficek$^{1,2}$\orcidlink{0000-0001-5885-7064} and Maciej Maliborski$^{1,2,3}$\orcidlink{0000-0002-8621-9761}}

\affil{$^1$University of Vienna, Faculty of Mathematics, Oskar-Morgenstern-Platz 1, 1090 Vienna, Austria}
\affil{$^2$University of Vienna, Gravitational Physics, Boltzmanngasse 5, 1090 Vienna, Austria}
\affil{$^3$TU Wien, Institute of Analysis and Scientific Computing, Wiedner Hauptstraße 8–10, 1040 Vienna, Austria}

\email{maciej.maliborski@tuwien.ac.at}

\begin{abstract}
As a first step in exploring time-periodic solutions of the Einstein equations with a negative cosmological constant, we study the cubic conformal wave equation on the Einstein cylinder. Using a combination of numerical and perturbative techniques, we discover that time-periodic solutions form intricate bifurcation patterns.
\end{abstract}

\section{Introduction}
\label{sec:Introduction}

\subsection{Motivation}
\label{sec:Motivation}

Motivated by the rich dynamics of asymptotically anti–de Sitter (AdS) spacetimes and, more generally, of confined Hamiltonian systems, our aim is to understand the dichotomy between stable and unstable scenarios. To begin, we explore time-periodic solutions with the intention to examine their impact on nonlinear dynamics. To this end, we study a simple model: the conformally invariant wave equation on the Einstein cylinder, which is especially attractive because of its conformal invariance and the simple form it takes under spherical symmetry \cite{Ficek.20254w4}.

Let us recall that the AdS space in four spacetime dimensions 
\begin{equation}
    \label{eq:25.08.25_01}
    \tilde{g} = \frac{\ell^{2}}{\cos^{2}{x}}\left(-\diff{t}^{2} + \diff{x}^{2} +\sin^{2}{x}\,\diff{\Omega}^{2}\right)\,,
    \quad
    x\in[0,\pi/2]\,.
\end{equation}
solves $\textrm{Ric}(\tilde{g})=-\frac{6}{\ell^{2}}\tilde{g}$. The main feature of this solution is the spatial infinity at $x = \pi/2$, which forms a timelike cylinder $\mathbb{R}\times\mathbb{S}^{2}$ with boundary metric $\bar{g}=-\diff{t}^{2}+\diff{\Omega}^{2}=-\diff{t}^{2}+\diff{\theta}^{2}+\sin^{2}{\theta}\,\diff{\varphi}^{2}$.

AdS spacetime is not globally hyperbolic: null geodesics reach the boundary in finite coordinate time. To obtain a well-posed dynamics one must therefore prescribe boundary conditions (BCs) at $x = \pi/2$. Once a choice is made, a natural question arises: do small perturbations of AdS remain small for all future times? For de Sitter and Minkowski spacetimes the answer is affirmative \cite{Friedrich.1986, ChristodoulouKlainerman+1994}. In contrast, in AdS the main stabilizing mechanism of Minkowski space---energy dispersion to infinity---may be absent, raising the possibility of very different dynamical behaviour.

A substantial body of work now points to the instability of AdS under reflective boundary conditions. The first numerical study of the spherically symmetric Einstein–scalar field system \cite{Bizoń.2011, Jałmużna.2011} showed that generic finite-amplitude $\varepsilon$ perturbations lead to black hole (BH) formation on timescales $\varepsilon^{-2}$, later confirmed in \cite{Buchel.2012}; see also \cite{Evnin.2021, Martinon.2017ve} for a detailed historical perspective. In addition, \cite{Bizoń.2011} conjectured the instability mechanism: turbulent energy transfer among linear modes, concentrating energy on sufficiently small scales to form trapped surfaces. Follow-up works developed perturbative expansions in $\varepsilon$, yielding effective equations for resonant mode interactions \cite{Craps.2014, Craps.2015, Green.2015}. These were later shown to develop an oscillatory singularity in finite time \cite{Bizoń.2015}, apparently corresponding to BH formation in the full Einstein–scalar field system. On the rigorous side, \cite{Moschidis.2023} proved the instability of AdS for the Einstein–massless Vlasov system, via an analysis of particle-beam interactions.

At the same time, evidence has emerged that turbulence may be absent for certain initial data, notably those close to time-periodic solutions. The first construction of asymptotically AdS time-periodic solutions in the Einstein–scalar field model of \cite{Bizoń.2011, Jałmużna.2011} was given in \cite{Maliborski.2013}, see also \cite{Maliborski.2015}. These works initiated many extensions, covering massive \cite{Kim.2015} and self-interacting fields \cite{Fodor.2014, Fodor.2015}, complex fields (boson stars) \cite{Buchel.2013}, the vacuum \cite{Dias.2012, Horowitz.2015, Dias.201861h, Dias.2016qnd, Fodor.2017, Martinon.2017}, and forced scalar fields at the conformal boundary \cite{Biasi.2018}.

Motivated by our recent results for the cubic wave equation on an interval with Dirichlet boundary conditions \cite{Ficek.2024A, Ficek.2024B, Ficek.20259zq}, we believe the findings of \cite{Maliborski.2013} should be revisited. Given the complexity of the Einstein equations, the full problem is expected to be even more intricate than in \cite{Ficek.2024A, Ficek.2024B, Ficek.20259zq}. As a first step, we therefore turn to a simpler toy model \cite{Ficek.20254w4}.

\subsection{Model}
\label{sec:Model}

We consider the conformally invariant cubic wave equation
\begin{equation}
    \label{eq:25.08.25_02}
    g^{\mu\nu}\nabla_{\mu}\nabla_{\nu}\phi - \frac{1}{6} R(g)\phi - \lambda\phi^3 = 0, \quad \lambda> 0\,,
\end{equation}
on the Einstein cylinder $\mathbb{R} \times \mathbb{S}^3$ with metric
\begin{equation}
    \label{eq:25.08.25_03}
    g = -\diff{t}^2 + \rho^2 \left( \diff{x}^2 + \sin^{2}{x} \,\diff{\Omega}^2 \right)\,,
\end{equation}
where $\rho$ is the radius of $\mathbb{S}^3$, $\diff\Omega^2$ is the round metric on $\mathbb{S}^2$, and $R(g) = 6/\rho^2$ is the Ricci scalar.

Assuming spherical symmetry $\phi=\phi(t,x)$ and setting $\phi(t,x) = u(t,x)/\sin x$, followed by the rescalings $t \to t/\rho$ and $\phi \to \rho \sqrt{\lambda},\phi$, we obtain a 1D wave equation with Dirichlet BCs (enforced by regularity of $\phi$ on $\mathbb{S}^3$):
\begin{equation}
    \label{eq:25.08.25_04}
    \partial_{t}^{2}u - \partial_{x}^{2}u + \frac{u^{3}}{\sin^{2}{x}} = 0\,,
    \quad
    u(t,0) = 0 = u(t,\pi)\,.
\end{equation}

A few remarks are in order:
\begin{itemize}
\item The conserved energy of \eqref{eq:25.08.25_04} is
 \begin{equation}
     \label{eq:25.08.25_05}
     E[u] = \int_{0}^{\pi}\left(\frac{1}{2}\left(\partial_{t}u\right)^{2}+\frac{1}{2}\left(\partial_{x}u\right)^{2}+\frac{1}{4}\frac{u^{4}}{\sin^{2}{x}}\right)\diff{x}\,.
 \end{equation}

\item By conformal symmetry between \eqref{eq:25.08.25_01} and \eqref{eq:25.08.25_02}, and using $\tilde{\phi}(t,x)=\phi(t,x)\cos{x}$, our solutions---being either symmetric or antisymmetric across the equator of $\mathbb{S}^{3}$---correspond to solutions of \eqref{eq:25.08.25_02} on AdS \eqref{eq:25.08.25_01}, with Dirichlet $\phi(t,\pi/2)=0$ or Neumann $\partial_{x}\phi(t,\pi/2)=0$ BCs. This link is noteworthy since nonlinear dynamics in AdS has been extensively studied in recent years \cite{Bizoń.2017ndf, Bizoń.2019, Bizoń.2020qd, Evnin.2021}, and our objectives closely align with that program.

\item In this work we focus on the defocusing case, leaving the qualitatively different focusing case for future study.
\end{itemize}

\subsection{Related studies}
\label{sec:RelatedStudies}

Time-periodic solutions of \eqref{eq:25.08.25_04} have been studied extensively. Rigorous results establish small-amplitude families bifurcating from single linear modes $\sin(nx)$, $n\in\mathbb{N}$, whose frequencies form Cantor-type sets \cite{Bizon.2017, Chatzikaleas.2020oh, Berti.2024, Silimbani.2024, Chatzikaleas.2024}.
Similar results hold for the cubic wave equation with Dirichlet boundary conditions (i.e. \eqref{eq:25.08.25_04} without the $\sin^{2}x$ factor), where the existence of such solutions was shown in \cite{Bambusi.2001, Berti.2003, Berti.2004}; see also the review \cite{berti2007nonlinear}.
The appearance of Cantor sets in these results arises from frequency restrictions imposed by certain Diophantine conditions, a necessary constraint to overcome the small divisor problem, yielding uncountable sets that accumulate at one with measure zero.
Recent studies, however, uncovered much richer large-amplitude structures \cite{Ficek.2024A, Ficek.2024B}. In this work, we explore these intricate patterns in the geometrically motivated model \eqref{eq:25.08.25_04}, thereby extending previous existence results into the finite-amplitude regime.

\section{Results}
\label{sec:Results}

In this section we summarize the main findings of \cite{Ficek.20254w4}, which include:
\begin{itemize}
    \item Identification of new classes of multi–mode, large–energy solutions,
    \item Revelation of a fractal–like network of trunks and branches, suggesting an infinite web of interconnected solutions,
    \item Proof that the Poincar\'{e}–Lindstedt (PL) perturbative series can be continued to arbitrary order,
    \item Demonstration that, once resummed by Pad\'{e} approximation, the PL series encodes detailed information about the complex bifurcation structure of time–periodic solutions.
\end{itemize}

\subsection{Galerkin approximation and complex structure}
\label{sec:GalerkinApproximationAndComplexStructure}

We rescale time $\tau = \Omega t$ so that the period is $2\pi$, $u(\tau,x) = u(\tau+2\pi,x)$, and the equation becomes
\begin{align}
    \label{eq:25.08.25_06}
    \Omega^2 \partial_\tau^2 u - \partial_x^2 u + \frac{u^3}{\sin^2 x} = 0\,.
\end{align}
Approximating solutions by finite Fourier series\footnote{Here we assume $N$ is even.}
\begin{equation}
	\label{eq:25.08.25_07} 
	u_{M}(\tau,x) = \sum_{j=0}^{M-1}\sum_{k=0}^{M-1}\hat{u}_{jk}\cos(2j+1)\tau\, \sin2(k+1)x ,
\end{equation}
and applying the Galerkin method yields an algebraic system for the coefficients $\hat{u}_{jk}$, solved via Newton iteration and continuation techniques \cite{Keller.1987, Allgower.1990}. This reveals that from each eigenmode $\sin Nx$, $N>1$ a complex network of branches emerges from the main part that we refer to as the trunk.\footnote{For $N = 1$, only the trunk exists and can be expressed exactly using Jacobi elliptic functions.} These branches densely populate the trunk, connect to rescaled versions of trunks from other eigenmodes, form an intricate pattern in the energy-frequency diagram, and become increasingly complex as the truncation order $M$ increases. Moreover, the existence of branches correlates with Cantor sets appearing in rigorous existence proofs \cite{Berti.2024,Silimbani.2024, Chatzikaleas.2020oh, Chatzikaleas.2024}.

To interpret this structure, we introduced reducible Galerkin systems \cite{Ficek.2024B}, finite-dimensional models retaining only a few resonant modes that couple in a minimal way. Their algebraic form makes them explicitly solvable, and their solutions reproduce the same trunk–branch bifurcations and rescalings. These simplified systems suggest that the observed complexity originates from resonant interactions of a small number of modes.

This naturally leads to the question: can perturbation theory, extended to sufficiently high order, capture such global features?

\subsection{Poincar\'{e}–Lindstedt construction and Pad\'{e} analysis}
\label{sec:PoincareLindstedtConstructionAndPadeAnalysis}

Time-periodic solutions are constructed via the PL method. Rescaling $u \to \sqrt{\varepsilon} u$, equation~\eqref{eq:25.08.25_06} becomes
\begin{align}
    \label{eq:25.08.25_08}
    \Omega^2 \partial_\tau^2 u - \partial_x^2 u + \varepsilon\frac{u^3}{\sin^2 x} = 0\,.
\end{align}
We expand both the frequency and the solution as formal series in $\varepsilon$:
\begin{align}
    \label{eq:25.08.25_09}
    \Omega^2 = N^2 + \sum_{n=1}^\infty \varepsilon^n \omega^{(n)}, 
    \qquad
    u(\tau, x) = \sum_{n=0}^\infty \varepsilon^n u^{(n)}(\tau,x)\,,
\end{align}
with $N \geq 1$, where $u^{(n)}$ are $2\pi$–periodic in $\tau$ and satisfy Dirichlet boundary conditions in $x$.  
Substitution into \eqref{eq:25.08.25_06} yields a hierarchy of equations ordered in $\varepsilon$, with the lowest order solved by $u^{(0)} = \cos{\tau}\sin{Nx}$.

At each order, resonant terms may arise that could obstruct continuation of the series.  
A central result of our work is that all resonances can be removed systematically, showing that the PL series extends to arbitrary order.\footnote{This is possible because, in the conformal case, the product $\sin jx\, \sin kx\, \sin lx / \sin^2 x$ generates harmonics only up to $\sin(j+k+l-2)x$, unlike the simple cubic case \cite{Ficek.2024A}.}  
This establishes the existence of formal solution families (labeled by $N$). To reach very high orders, projection integrals were evaluated numerically in extended precision (details are given in \cite{Ficek.20254w4}).

Although formal, the PL expansion encodes information about the global solution structure. To extract it, we apply Pad\'{e} approximation \cite{Baker.1996} to both the frequency series and Fourier coefficients. Writing \eqref{eq:25.08.25_09} as
\begin{align}
    \label{eq:25.08.25_10}
    u(\tau,x) = \sum_{m=0}^\infty a_{2m+1}(\varepsilon) \cos(2m+1)\tau\,\sin (2m+1)Nx 
    + \sum_{j=0}^\infty \sideset{}{'}\sum_{k=1}^\infty b_{2j+1,k}(\varepsilon) \cos(2j+1)\tau\,\sin kx\,,
\end{align}
we construct diagonal Pad\'{e} approximants $[n/n]$ of coefficients, e.g.
\begin{equation}
    \label{eq:25.08.25_11}
    a_{1}(\varepsilon) = a_{1}^{(0)} + a_{1}^{(1)}\varepsilon + \cdots + a_{1}^{(2n)}\varepsilon^{2n}
    \approx [n/n] = \frac{p_{0} + p_{1}\varepsilon + \cdots + p_{n}\varepsilon^{n}}{1 + q_{1}\varepsilon + \cdots + q_{n}\varepsilon^{n}}\,.
\end{equation}
Plotting the parametric curve $(|a_{1}|,\Omega)$ shows that Pad\'{e} approximants reproduce the main trunk far beyond the radius of convergence of the formal series. Moreover, the poles of the approximants align with the frequencies of bifurcating branches.

Our procedure for locating branches is:
\begin{enumerate}
    \item Select a Fourier coefficient in \eqref{eq:25.08.25_10} and compute its $[n/n]$ Pad\'{e} approximant.
    \item Find real roots $\varepsilon_{*}$ of the denominator.
    \item Evaluate $\Omega(\varepsilon_{*})$ from its Pad\'{e} approximant.
    \item Track these values as $n$ increases.
\end{enumerate}
The poles cluster in patterns reflecting the bifurcation structure.  
Comparison with reducible Galerkin predictions confirms that Pad\'{e} analysis not only reproduces the main trunk but also encodes the location and accumulation of branches, in agreement with numerical studies. Thus Pad\'{e} analysis provides an independent confirmation of the intricate solution structure revealed by Galerkin methods.

\section*{Acknowledgements}
We acknowledge the support of the Austrian Science Fund (FWF) through Project \href{http://doi.org/10.55776/P36455}{P 36455}, Wittgenstein Award \href{http://doi.org/10.55776/Z387}{Z 387}, and the START-Project \href{http://doi.org/10.55776/Y963}{Y 963}.Computations were performed on a supercomputer hosted by the Institute of Theoretical Physics at Jagiellonian University, with funding provided by the Polish National Science Centre grant No. 2017/26/A/ST2/00530.

\bibliography{lib}

\end{document}